\documentclass[%
aip,
jcp,%
amsmath,amssymb,
reprint%
]{revtex4-1}
\usepackage[latin1]{inputenc}

\usepackage{amsmath}
\usepackage{amsfonts}
\usepackage{amssymb,bm}
\usepackage{graphicx}
\usepackage{physics}
\usepackage{upgreek}
\usepackage[outercaption]{sidecap}   
\usepackage{color}
\usepackage[normalem]{ulem}
\mathchardef\mhyphen="2D
\usepackage{txfonts}

\DeclareMathAlphabet{\pazocal}{OMS}{zplm}{m}{n}

\begin{document}

\def\bra#1{\left<{#1}\right|}
\def\ket#1{\left|{#1}\right>}
\def\expval#1#2{\bra{#2} {#1} \ket{#2}}
\def\mapright#1{\smash{\mathop{\longrightarrow}\limits^{_{_{\phantom{X}}}{#1}_{_{\phantom{X}}}}}}

\title{Confirming the role of nuclear tunnelling in aqueous ferrous-ferric electron transfer}
\author{Joseph E. Lawrence}
\email{joseph.lawrence@chem.ox.ac.uk}
\affiliation{Department of Chemistry, University of Oxford, Physical and Theoretical\\ Chemistry Laboratory, South Parks Road, Oxford, OX1 3QZ, UK}
\author{David E. Manolopoulos}
\affiliation{Department of Chemistry, University of Oxford, Physical and Theoretical\\ Chemistry Laboratory, South Parks Road, Oxford, OX1 3QZ, UK}

\begin{abstract}
We revisit the well-known aqueous ferrous-ferric electron transfer reaction in order to address recent  suggestions that nuclear tunnelling can lead to significant deviation from the linear response assumption inherent in the Marcus picture of electron transfer. A recent study of this reaction by Richardson and coworkers has found a large difference between their new path-integral method, GR-QTST, and the saddle point approximation of Wolynes (Wolynes theory). They suggested that this difference could be attributed to the existence of multiple tunnelling pathways, leading Wolynes theory to significantly overestimate the rate. This was used to argue that the linear response assumptions of Marcus theory may break down for liquid systems when tunnelling is important. If true, this would imply that the commonly used method for studying such systems, where the problem is mapped onto a spin-boson model, is invalid. However, we have recently shown that size inconsistency in GR-QTST can lead to poor predictions of the rate in systems with many degrees of freedom. We have also suggested an improved method, the path-integral linear golden-rule (LGR) approximation, which fixes this problem. Here we demonstrate that the GR-QTST results for ferrous-ferric electron transfer are indeed dominated by its size consistency error. Furthermore, by comparing the LGR and Wolynes theory results, we confirm the established picture of nuclear tunnelling in this system. Finally, by comparing our path-integral results to those obtained by mapping onto the spin-boson model, we reassess the importance of anharmonic effects and the accuracy of this commonly used mapping approach.
\end{abstract}

\maketitle
\section{Introduction}
The most commonly used approach to understand electron electron transfer, in biological, inorganic and materials chemistry, is Marcus theory.\cite{Balzani96,Holm96,Bredas04,Kamat08} In its basic form, the Marcus picture of electron transfer consists of two steps: spontaneous thermal fluctuations of the solvent polarisation (assumed to be harmonic) first bringing the reactant and product charge transfer states to the same energy, followed by a golden-rule transition from one charge transfer state to the other.\cite{Marcus56,Marcus85,Hush61,Hush99} While there are many analytical theories which go beyond the simplest form of Marcus theory, for example by including nuclear quantum effects,\cite{Levich59,Ovchinnikov69,Ulstrup79,Dogonadze80,Weiss08,Kestner74,Song93a} the majority still assume that the relaxation of the solvent polarisation (or diabatic energy gap) can be treated within linear response theory, which is equivalent to assuming a harmonic (spin-boson) model for the diabatic potentials.\cite{Georgievskii99} 

Perhaps the prototypical example of electron transfer, commonly used to introduce Marcus theory in undergraduate text books,\cite{Henderson93,Weller18} is the aqueous ferrous-ferric electron transfer reaction,
\begin{equation}
\mathrm{Fe}^{2+}{\rm (aq)}+ \mathrm{Fe}^{3+}{\rm (aq)}\rightarrow\mathrm{Fe}^{3+}{\rm (aq)}+\mathrm{Fe}^{2+}{\rm (aq)},
\end{equation}
which, along with other similar self-exchange reactions, has been the subject of numerous theoretical studies.\cite{Brunschwig80,Siders81,Kuharski88,Bader89,Bader90,Marchi93,Song93b,Ando01,Blumberger05,Blumberger06a,Blumberger08,Drechsel-Graua12} 
 In the classical limit the pioneering atomistic studies of ferrous-ferric electron transfer by Chandler and coworkers demonstrated that the assumptions of Marcus theory are well founded.\cite{Kuharski88,Bader89} By calculating the free-energy as a function of solvent polarisation they showed that the resulting surfaces were essentially parabolic in line with the Marcus assumption,\cite{Kuharski88} and by studying the solvent relaxation immediately after electron transfer they were able to illustrate the close agreement between the exact and linear response results.\cite{Bader89}

In addition to investigating the ferrous-ferric system in the classical limit, Chandler and coworkers also carried out one of the first atomistic studies of the influence nuclear tunnelling on electron transfer.\cite{Bader90} In this work, they used imaginary-time path-integral simulations to directly calculate the Wolynes theory\cite{Wolynes87} approximation to the electron transfer rate, as well as performing classical equilibrium simulations to map the problem onto the spin-boson model. The Wolynes theory approximation to the quantum rate can be understood in terms of making a Gaussian (saddle point) approximation to the exact reactive flux autocorrelation function.\cite{Wolynes87} As it requires no real time information it can easily be applied to fully atomistic models of electron transfer,\cite{Zheng89,Zheng91} with no additional approximations, by exploiting the well-known quantum classical isomorphism.\cite{Chandler81,Parrinello84} An alternative approach to calculating the influence of nuclear tunnelling on the electron transfer rate is to map the problem onto the spin-boson model, for which there exists a simple closed form expression for the exact quantum mechanical golden-rule rate.\cite{Nitzan06} Mapping to the spin-boson Hamiltonian is equivalent to assuming the relaxation of the solvent polarisation can be described via linear response. It provides a simple way to estimate nuclear quantum effects from classical atomistic simulations, by calculating the spectral density from the classical energy gap autocorrelation function in the reactant ensemble.\cite{Warshel86,Hwang97}  This approach has since become a popular way to use atomistic simulation to study electron transfer, finding application in a diverse range of problems from electron transfer in enzymes to device physics.\cite{Ando97,Ungar99,Ando01,Tipmanee10,Blumberger15,Firmino16,Tong20} Using both methods Chandler and coworkers found surprisingly large tunnelling enhancements of the ferrous-ferric electron transfer rate, with Wolynes theory predicting a quantum enhancement factor of $\sim$60 and the spin-boson mapping a factor of $\sim$40 at room temperature.\cite{Bader90} These large quantum enhancements proved quite controversial and have since been the focus of several papers.\cite{Marchi93,Song93b,Blumberger08}  The general consensus that has emerged from these papers is that the large enhancement is due to a lack of polarisability in the SPC water model that Chandler and coworkers used in their calculations.\cite{Bader90} Unfortunately, unlike in the classical limit where the approximations of Marcus theory can be validated using the exact classical golden-rule rate, in the quantum regime it is not possible to calculate the exact quantum rate for anything beyond simple model systems and hence the accuracy of Wolynes theory and mapping onto the spin-boson model can only be inferred indirectly.

Over the past 30 years there has been a steady development of imaginary-time path-integral techniques, and for systems obeying the Born-Oppenheimer approximation they have become a routine way to study the influence of nuclear quantum effects in a wide range of chemical systems.\cite{Voth89,Cao94a,Cao94b,Cao94c,Cao94d,Miller03,Craig04,Vanicek05,Craig05a,Craig05b,Collepardo08,Boekelheide11,Habershon13,Richardson09,Hele15} In the last few years this has led to a renewed interest in the use of path-integral approaches to study electronically non-adiabatic problems such as electron transfer,\cite{Shushkov12,Richardson13,Ananth13,Duke16,Menzeleev14,Kretchmer16,Chowdhury17,Kretchmer18,Tao18,Tao19,Lawrence18,Lawrence19b,Lawrence19a,Lawrence20a,Lawrence20b,Richardson15a,Richardson15b,Heller20,Thapa19,Fang19,Fang20} building on the early work by Wolynes, Chandler and others.\cite{Wolynes87,Zheng89,Zheng91,Cao95,Cao97,Schwieters98,Schwieters99} Recently, Richardson and coworkers have reassessed the accuracy of Wolynes theory, and have pointed out that its Gaussian (saddle point) approximation can breakdown severely for systems with multiple transition states.\cite{Fang19} As an alternative to Wolynes theory, they have suggested a new path-integral based approach, which they call golden-rule quantum transition state theory (GR-QTST).\cite{Thapa19,Fang19} They have shown that, for low-dimensional model systems designed to exhibit multiple transition states as well as significant nuclear quantum effects, GR-QTST gives accurate rate predictions, while Wolynes theory can overestimate the rate by more than two orders of magnitude.\cite{Fang19} On applying GR-QTST to a fully atomistic model of ferrous-ferric electron transfer,  they found a tunnelling enhancement around 6 times smaller than that of Wolynes theory, and used this to argue that Wolynes theory may be overestimating the rate due to the presence of multiple tunnelling pathways.\cite{Fang20} This was then used to argue that even the linear response assumption may break down in the presence of tunnelling for ferrous-ferric electron transfer, casting doubt not only on the quantitative accuracy but also the qualitative validity of mapping to the spin-boson model.\cite{Fang20}  However, as was acknowledged by Richardson and coworkers, they were unable to conclusively prove that it was Wolynes theory that was breaking down for this system, and hence they could not rule out an error on the part of GR-QTST. 

In a recent paper,\cite{Lawrence20c} we have demonstrated that GR-QTST can suffer from a serious size inconsistency issue for large systems.  Here we argue that it is this size inconsistency that is responsible for the large difference between the Wolynes and GR-QTST predictions of the rate in ferrous-ferric electron transfer, rather than a breakdown of Wolynes theory. To support this argument, and help assess both the accuracy of Wolynes theory and the spin-boson mapping for this system, we also apply an alternative path-integral method, the linear golden-rule (LGR) approximation.\cite{Lawrence20c} Since it is both size consistent and does not assume the correlation function is Gaussian, LGR provides a viable alternative to Wolynes theory and the spin-boson mapping method for including nuclear quantum effects in condensed phase electron transfer reactions. 

Section \ref{Theory_Section} gives an overview of the relevant electron transfer theory. Sec.~\ref{Computational_Section} provides some details of our simulations of the ferrous-ferric system. Sec.~\ref{Results_Section1} analyses how a lack of size consistency in GR-QTST affects its prediction of the ferrous-ferric rate, and investigates the recent suggestion that Wolynes theory is breaking down for this reaction.\cite{Fang20} Sec.~\ref{Results_Section2} reassesses the validity of the spin-boson mapping in the light of the preceding sections, and Sec.~\ref{Conclusion} concludes the paper.

\section{Theory} \label{Theory_Section}
\subsection{Exact theory}
The standard Hamiltonian for describing electron transfer can be written in the form 
\begin{equation}
\hat{H}=\hat{H}_0\dyad{0}{0}+\hat{H}_1\dyad{1}{1} + \Delta\big(\dyad{0}{1}+\dyad{1}{0}\big)
\end{equation}
where
\begin{equation}
\hat{H}_i = \sum_{\nu=1}^{f} \frac{\hat{p}^2_\nu}{2m_\nu} + \hat{V}_i(\bm{q})
\end{equation}
is the diabatic Hamiltonian for state $i$, with potential $\hat{V}_i(\bm{q})$,  and $\Delta$ is the electronic coupling. We follow previous authors and make the Condon approximation, assuming that $\Delta$ is independent of the nuclear coordinates.\cite{Bader90,Fang20} The two diabatic electronic states $\ket{0}$ and $\ket{1}$ correspond to the two possible charge transfer states, e.g.~Fe(II)-Fe(III) and Fe(III)-Fe(II), for which the potential energy surfaces cross. 

The coupling between the two diabatic states is usually assumed to be sufficiently small that the rate to go from one charge transfer state to the other is accurately described by Fermi's golden rule. The rate can then be written in terms of the time integral of a modified flux-flux correlation function as\cite{Lax52,Kubo55,Kestner74,Wolynes87}
\begin{equation}
k = \frac{\Delta^2}{Q_r\hbar^2} \int_{-\infty}^{\infty} c_{\lambda}(t)\, \mathrm{d}t \label{exact_rate}
\end{equation}
where $Q_r=\tr[e^{-\beta\hat{H}_0}]$ is the reactant partition function and the correlation function is given by
\begin{equation}
c_{\lambda}(t)= c(t+i\lambda\hbar) = \tr[e^{-(\beta-\lambda)\hat{H}_0}e^{-i\hat{H}_0 t/\hbar}e^{-\lambda\hat{H}_1}e^{+i \hat{H}_1 t/\hbar}]. \label{Correlation_Function}
\end{equation}
It is important to note that the exact golden-rule rate constant $k$ is independent of the shift in imaginary time $i\lambda \hbar$, which changes both the initial value of the correlation function and how rapidly it oscillates.

Alternatively we can rewrite Eq.~(\ref{exact_rate}) as
\begin{equation}
k = \frac{2\pi\Delta^2}{Q_r\hbar} \rho_{\lambda}(0) e^{-\beta F(\lambda)}, \label{epsilon_dep_rate}
\end{equation}
where $\rho_{\lambda}(E)$ is a probability distribution given by the Fourier transform 
\begin{equation}
\rho_{\lambda}(E) = \frac{1}{2\pi\hbar} \int_{-\infty}^{\infty} \Big\langle e^{-i\hat{H}_0 t/\hbar}e^{+i \hat{H}_1 t/\hbar} \Big\rangle_\lambda e^{-iE t/\hbar}  \mathrm{d}t. \label{Exact_distribution}
\end{equation}
Here,
\begin{equation}
\left\langle \hat{A} \right\rangle_\lambda = \frac{\tr\left[ e^{-\lambda \hat{H}_1} e^{-(\beta-\lambda)\hat{H}_0}  \hat{A} \right]}{\tr\left[ e^{-\lambda \hat{H}_1} e^{-(\beta-\lambda)\hat{H}_0}   \right]}
\end{equation}
such that 
\begin{equation}
\Big\langle e^{-i\hat{H}_0 t/\hbar}e^{+i \hat{H}_1 t/\hbar} \Big\rangle_\lambda = \frac{c_{\lambda}(t)}{c_{\lambda}(0)}
\end{equation}
is the correlation function normalised to one at $t=0$, and the Boltzmann factor,
\begin{equation}
e^{-\beta F(\lambda)}=c_{\lambda}(0)=c(i\lambda\hbar)=\tr[e^{-(\beta-\lambda)\hat{H}_0}e^{-\lambda \hat{H}_1}], \label{imaginary_cff}
\end{equation}
is the correlation function evaluated on the imaginary-time axis. Note that $e^{-\beta F(0)}={\rm tr}[e^{-\beta\hat{H}_0}]=Q_r$, so $F(0)$ is just the free energy of the reactants.

Unfortunately exact evaluation of $\rho_\lambda(E)$ for anything other than simple models is generally not possible, as it either requires a knowledge of all of the eigenstates of $\hat{H}_0$ and $\hat{H}_1$ or the numerical convergence of a real-time path integral, which exhibits a well-known sign problem. Despite the difficulty with real time, evaluation of the correlation function for purely imaginary time is straightforward even for arbitrarily complex systems.\cite{Wolynes87,Cao97,Lawrence18} This is because, as is clear from Eq.~(\ref{imaginary_cff}), the correlation function on the imaginary-time axis, $e^{-\beta F(\lambda)}$, can be thought of as a generalised quantum partition function, which can be written in terms of the partition function of a classical ring polymer by exploiting the well-known imaginary-time path-integral classical isomorphism.\cite{Chandler81,Parrinello84} This isomorphism makes the evaluation of quantities on the imaginary-time axis, such as the free energy difference $F(\lambda)-F(0)$, entirely straightforward, with a computational effort that scales only linearly with system size. But in order to calculate $\rho_{\lambda}(0)$ (and hence the rate) for a complex system such as ferrous-ferric electron transfer, one must make some kind of approximation. 

In this paper we will consider two different types of approximation. The first is to use an approximate expression for $\rho_{\lambda}(E)$ which can be directly evaluated using imaginary-time path-integral simulations of the fully atomistic model of the reaction. The methods in this category can be thought of as transition state theories, as they use only statistical information to approximate the rate. The second, and more commonly used, approach is to map the system onto a spin-boson model\cite{Warshel86,Bader90} by constructing a harmonic bath designed to mimic the influence of the nuclear coordinates on the electronic states. The quantum mechanical golden-rule rate for the model can then be calculated simply by using the exact closed form expression for $c(t)$.\cite{Levich59,Ovchinnikov69,Ulstrup79,Dogonadze80,Weiss08} Each of these approaches has its advantages and disadvantages, and as we shall demonstrate the accuracy of each approach is strongly dependent on the details of the method that is used. 

\subsection{Mapping to the spin-boson model}
The spin-boson model assumes that from the perspective of the electron transfer the nuclear potentials in both charge transfer states can be modelled as being purely harmonic with the same frequencies in both states, 
\begin{subequations}
\begin{align}
V_0(\bm{q}) &= \sum_{\nu=1}^f \frac{1}{2}m\omega_\nu^2 q_\nu^2 + c_\nu q_\nu \\
V_1(\bm{q}) &= \sum_{\nu=1}^f \frac{1}{2}m\omega_\nu^2 q_\nu^2 - c_\nu q_\nu -\epsilon,
\end{align} 
\label{SB_diabats}
\end{subequations}
where $\epsilon=F(0)-F(\beta)$ is the thermodynamic driving force for the reaction. The effect of the nuclear dynamics on the electronic coordinate is then completely described by the spectral density, which is formally defined as
\begin{equation}
J(\omega) = \frac{\pi}{2}\sum_{\nu=1}^f \frac{c_\nu^2}{m\omega_\nu} \delta(\omega-\omega_\nu),
\end{equation}
and the exact rate can be calculated using\cite{Levich59,Ovchinnikov69,Ulstrup79,Dogonadze80,Weiss08}
\begin{equation}
\frac{c(t)}{Q_r} = \exp(-i \epsilon t/\hbar - \phi(t)/\hbar),
\end{equation}
where
\begin{equation}
\phi(t) = \frac{4}{\pi}\int_{-\infty}^{\infty}\frac{J(\omega)}{\omega^2}\Bigg[\frac{1-\cos(\omega t)}{\tanh(\beta\hbar\omega/2)}-i \sin(\omega t)\Bigg]\mathrm{d}\omega.
\end{equation}

To map a real system onto the spin-boson model one must therefore choose a method for defining the spectral density of the system. There are many ways that this can be achieved, using either simulation or experimental data.\cite{Ulstrup79,Song93b,Gilmore08} The approach we will focus on here is the use of classical molecular dynamics to define the spectral density according to the relation\cite{Warshel86,Bader90,Blumberger15} 
\begin{equation}
\frac{J(\omega)}{\omega}=\frac{\Lambda}{2}\int_0^{\,\infty} \frac{\big\langle \delta V_{-}(0) \, \delta V_{-}(t)\big\rangle_{\mathrm{cl},0}}{\big\langle  \delta V_{-}^2(0)\big\rangle_{\mathrm{cl},0}}  \cos(\omega t) \mathrm{d}t. \label{Scaled_J_w}
\end{equation}
Here the averages are taken in the classical canonical ensemble of the reactants, $V_-(t)=V_0(t)-V_1(t)$ is the energy gap between the reactant and product diabats at time $t$ along a microcanonical trajectory sampled from this ensemble, and the energy gap fluctuation is defined as
\begin{equation}
\delta{V}_{-}(t) = {V}_{-}(t)-\langle {V}_{-}\rangle_{\mathrm{cl},0}.
\end{equation}
The Marcus theory reorganisation energy $\Lambda$ in Eq.~(\ref{Scaled_J_w}) is taken to be 
 \begin{equation}
 \Lambda = \epsilon - \langle {V}_{-}\rangle_{\mathrm{cl},0}, \label{Lambda_Egap}
 \end{equation}
 with the driving force given by 
 \begin{equation}
\epsilon = \frac{1}{\beta}\ln(\frac{\tr_{\mathrm{cl}}\big[e^{-\beta{H}_1}\big]}{\tr_{\mathrm{cl}}\big[e^{-\beta{H}_0}\big]}),
 \end{equation}
 which is clearly zero for the symmetric ferrous-ferric system we shall consider here.
  
 (We note in passing that an alternative choice of mapping gives the Marcus theory reorganisation energy as
\begin{equation}
\Lambda_2= \frac{\beta}{2}  \big\langle \delta V_{-}(0) \, \delta V_{-}(0)\big\rangle_{\mathrm{cl},0}. \label{Lambda_def_2}
\end{equation}
While this is equivalent to Eq.~(\ref{Lambda_Egap}) for the spin-boson model, the two expressions differ for more realistic (atomistic and anharmonic) models of electron transfer. For completeness Appendix \ref{Mapping_to_SB_apdx} gives a detailed discussion of the origin of this choice of mapping, including a discussion of why Eq.~(\ref{Lambda_Egap}) is to be preferred over Eq.~(\ref{Lambda_def_2}).

There are a number of obvious difficulties with trying to map a complex anharmonic system onto a model as simple as the spin-boson model. While it may be reasonable for a symmetric reaction such as ferrous-ferric electron transfer to treat the reactants and product states as having the same frequencies, it is not clear that this should be valid for asymmetric reactions. Furthermore it is not clear how well in general, even for a symmetric reaction such as ferrous-ferric electron transfer, the motion to the transition state can be described by a collection of small fluctuations from equilibrium, given that the underlying potential is anharmonic.  

\subsection{High temperature limit}
In the high temperature limit, the exponent of the spin-boson correlation function can be simplified to a quadratic function of time, 
\begin{equation}
\frac{c(t)}{Q_r} \simeq e^{-i \epsilon t/\hbar - \beta\Lambda\big(t^2/\beta^2\hbar^2-it/\beta\hbar\big)},
\end{equation}
corresponding to the quadratic free energy
\begin{equation}
-\beta [F(\lambda)-F(0)] = \lambda\epsilon+\beta\Lambda\bigg(\frac{\lambda^2}{\beta^2}-\frac{\lambda}{\beta}\bigg).
\end{equation}
Hence the golden-rule rate in Eq.~(\ref{exact_rate}) reduces to the famous Marcus expression\cite{Marcus56,Marcus85} 
\begin{equation}
k_{\mathrm{MT}}= \frac{\Delta^2}{\hbar}\sqrt{\frac{\pi\beta}{\Lambda}}e^{-\beta(\Lambda-\epsilon)^2/4\Lambda}. \label{Marcus_Theory}
\end{equation}

This expression has the appeal of being a simple analytic formula. However, if one is going to use a fully atomistic simulation to calculate the reorganisation energy, then calculating the high temperature limit of the exact rate is not significantly more difficult.\cite{Sun18} To see this, note that the high temperature limit of Eq.~(\ref{Exact_distribution}) is simply\cite{Chandler97,Sumi01}
\begin{equation}
\rho_{\mathrm{cl},\lambda}(E) = \big\langle\delta(V_{0}-V_{1}+E)\big\rangle_{\mathrm{cl},\lambda}, \label{Exact_High_Temp}
\end{equation}  
i.e.~the probability density along the diabatic energy gap coordinate, with the expectation value defined as
\begin{equation}
\langle A \rangle_{\mathrm{cl},\lambda} = \frac{\int \mathrm{d}^{f} \bm{p} \int \mathrm{d}^{f} \bm{q} \, A(\bm{q}) e^{-(\beta-\lambda)H_0(\bm{p},\bm{q})-\lambda H_1(\bm{p},\bm{q})}}{\int \mathrm{d}^{f} \bm{p} \int \mathrm{d}^{f} \bm{q}\, e^{-(\beta-\lambda)H_0(\bm{p},\bm{q})-\lambda H_1(\bm{p},\bm{q})}}.
\end{equation}
The high temperature limit of Eq.~(\ref{imaginary_cff}) is
\begin{equation}
e^{-\beta F_{\mathrm{cl}}(\lambda)} = \frac{1}{(2\pi\hbar)^{f}}\!\! \int \mathrm{d}^{f} \bm{p} \!\!\int \mathrm{d}^{f} \bm{q} e^{-(\beta-\lambda)H_0(\bm{p},\bm{q})-\lambda H_1(\bm{p},\bm{q})},\! \label{One_Bead_limit}
\end{equation}
from which it is clear that one can calculate the Boltzmann factor
\begin{equation}
{e^{-\beta F_{\rm cl}(\lambda)}\over Q_{r,{\rm cl}}} = e^{-\beta[F_{\rm cl}(\lambda)-F_{\rm cl}(0)]}
\end{equation}
with a thermodynamic integration. (For a symmetric electron transfer reaction such as the ferrous-ferric reaction, it is best to evaluate $\rho_{{\rm cl},\lambda}(0)$ and $e^{-\beta [F_{\rm cl}(\lambda)-F_{\rm cl}(0)]}$ at $\lambda=\beta/2$.) There is thus no difficulty in evaluating the exact golden-rule rate in Eq.~(\ref{epsilon_dep_rate}) in the high temperature limit, where it is given by a simple classical transition state theory:
\begin{equation}
k_{\rm cl} =  {2\pi\Delta^2\over\hbar}\left<\delta(V_0-V_1)\right>_{\rm cl,\lambda}e^{-\beta[F_{\rm cl}(\lambda)-F_{\rm cl}(0)]}. \label{Classical_Golden_Rule}
\end{equation}
Note that this avoids mapping the atomistic model onto a spin-boson model entirely.

\subsection{Wolynes theory}
For many electron transfer reactions it is well known that tunnelling and zero point energy can have a large effect on the rate, and so Marcus theory [Eq.~(\ref{Marcus_Theory})] and the classical golden-rule rate [Eq.~(\ref{Classical_Golden_Rule})] are not sufficient. While mapping to the spin-boson model has the promise of being able to capture nuclear quantum effects, as we have already discussed it is not clear that in this will always be a valid approach. An alternative way to include nuclear quantum effects is Wolynes theory,\cite{Wolynes87} which exploits the fact that it is straightforward to calculate the correlation function and its derivatives on the imaginary-time axis and use them to construct a short time (saddle point) approximation to the rate.

Wolynes suggested truncating a cumulant expansion of the exact correlation function in Eq.~(\ref{Correlation_Function}),
\begin{equation}
 \ln(\frac{c_{\lambda}(t)}{c_{\lambda}(0)}) =  i\mu_{1,\lambda}t-\mu_{2,\lambda}t^2/2 + \dots,
\end{equation}
at the second cumulant to give the Gaussian approximation
\begin{equation}
c_{\lambda}(t)\simeq c_{\mathrm{WT},\lambda}(t) = c_{\lambda}(0) e^{+i\mu_{1,\lambda}t-\mu_{2,\lambda}t^2/2} \label{Wolynes_Correlation}
\end{equation}
where since the derivatives of the correlation function with respect to real time are equivalent to those with respect to imaginary time, we have $\mu_{1,\lambda}=\beta F'(\lambda)$ and $\mu_{2,\lambda}=-\beta F''(\lambda)$. While the exact rate is independent of the value of $\lambda$ at which it is evaluated, the second order cumulant approximation to the rate is strongly dependent on the choice of $\lambda$, and it is most accurate at the saddle point of the correlation function where
\begin{equation}
 \mu_{1,\lambda_{\mathrm{sp}}} \equiv \beta F'(\lambda_{\mathrm{sp}})=0. \label{Saddle_Point_Condition}
 \end{equation}
Substituting Eq.~(\ref{Wolynes_Correlation}) into Eq.~(\ref{exact_rate}) and evaluating the time integral with this value of $\lambda$ gives the Wolynes theory approximation to the rate\cite{Wolynes87}
\begin{equation}
k_{\mathrm{WT}}=\frac{\Delta^2}{Q_r\hbar}\sqrt{\frac{2\pi}{-\beta F''(\lambda_{\mathrm{sp}})}}e^{-\beta F(\lambda_{\mathrm{sp}})}.
\end{equation}

Alternatively we can think of Wolynes theory as making a Gaussian approximation to the exact distribution $\rho_\lambda(E)$ in Eq.~(\ref{Exact_distribution}),
\begin{equation}
\rho_{\mathrm{WT},\lambda}(E) = \sqrt{\frac{1}{2\pi\mu_{2,\lambda}}} \exp(-\frac{(E-\mu_{1,\lambda})^2}{2\mu_{2,\lambda}}),
\end{equation}
with the first two moments chosen to match those of the exact distribution (see Appendix \ref{Moments_Appendix})
\begin{equation}
\mu_{1,\lambda} = \int_{-\infty}^{\infty} E \rho_{\lambda}(E) \mathrm{d}E = \beta F'(\lambda)
\end{equation}
and
\begin{equation}
\mu_{2,\lambda} = \int_{-\infty}^{\infty} (E^2-\mu_{1,\lambda}^2) \rho_{\lambda}(E) \mathrm{d}E = -\beta F''(\lambda). \label{Second_moment}
\end{equation}
Viewed from this perspective, we see that the saddle point condition is equivalent to choosing the distribution for which the mean is at zero, $\mu_{1,\lambda}=0$. Provided the distribution is singly peaked, the approximation to $\rho_{\lambda}(E)$ is likely to be most accurate near its mean $\mu_{1,\lambda}$.  Hence, as it is $\rho_{\lambda}(0)$ that enters the expression for the rate constant in Eq.~(\ref{epsilon_dep_rate}), the saddle point condition gives the most accurate Wolynes theory rate.

All of the quantities involved in the Wolynes expression for the rate they can be calculated exactly using standard imaginary-time path-integral techniques. These path-integral calculations are straightforward to perform using classical molecular dynamics, by exploiting the well known quantum-classical ``isomorphism''. For a system with $f$ nuclear degrees of freedom, the correlation function can be expressed as the classical partition function of an $n$ bead ring polymer,
\begin{equation}
e^{-\beta F(\lambda)} \simeq \frac{1}{(2\pi\hbar)^{nf}} \int \mathrm{d}^{nf}\mathbf{q} \int \mathrm{d}^{nf}\mathbf{p} \, e^{-\beta_n h_n(\mathbf{p},\mathbf{q})-\beta_n U_n^{(l)}(\mathbf{q})}
\end{equation}
where $\beta_n=\beta/n$ and  $l/n=\lambda/\beta$, with the equality becoming exact in the $n\to\infty$ limit. Here $h_n(\mathbf{p},\mathbf{q})$ is the well known free ring-polymer Hamiltonian
\begin{equation}
h_{n}(\mathbf{p},\mathbf{q})= \sum_{j=1}^{n}\sum_{\nu=1}^f \bigg[ \frac{p_{j,\nu}^2}{2 m_{\nu}} + \frac{1}{2} m_\nu \omega_n^2\big(q_{j,\nu}-q_{j-1,\nu}\big)^2\bigg],
\end{equation}
with $\omega_n=1/\beta_n\hbar$ and $q_{0,\nu}\equiv q_{n,\nu}$, and the external potential is given by
\begin{equation}
U^{(l)}_{n}(\mathbf{q})\! =  \sum_{j=0}^{l-1}\frac{V_{1}(\bm{q}_j)\!+\!V_{1}(\bm{q}_{j+1})}{2} + \sum_{j=l}^{n-1} \frac{V_{0}(\bm{q}_j)\!+\!V_{0}(\bm{q}_{j+1})}{2},
\end{equation}
such that beads $\boldsymbol{q}_1\dots \boldsymbol{q}_{l-1}$ experience the product diabatic potential, beads $\boldsymbol{q}_{l+1} \dots \boldsymbol{q}_{n-1}$ experience the reactant potential, and beads $\boldsymbol{q}_{0}$ and $\boldsymbol{q}_{l}$ are ``bridging beads'' which experience the average diabatic potential. All of the quantities needed to evaluate the Wolynes rate, as well as those in GR-QTST and LGR discussed below can then simply be calculated using expectation values in the corresponding ensemble,
\begin{equation}
\left\langle A(\mathbf{q}) \right\rangle_\lambda = \lim_{n\to\infty} \frac{\int \mathrm{d}^{nf}\mathbf{q} \int \mathrm{d}^{nf}\mathbf{p} \, e^{-\beta_n h_n(\mathbf{p},\mathbf{q})-\beta_n U_n^{(l)}(\mathbf{q})} A(\mathbf{q}) }{\int \mathrm{d}^{nf}\mathbf{q} \int \mathrm{d}^{nf}\mathbf{p} \, e^{-\beta_n h_n(\mathbf{p},\mathbf{q})-\beta_n U_n^{(l)}(\mathbf{q})}}.
\end{equation}

Although Wolynes theory is very accurate for many systems, including the spin-boson model, recent work by Richardson and coworkers has demonstrated that it can break down for systems with multiple transition states.\cite{Fang19} This can be understood by considering a reaction which can be separated into two distinct reaction pathways $a$ and $b$, such that the rate can be written as a sum of contributions from each, $k = k_a + k_b$. For such a system it follows that the distribution can be written as $\rho_{\lambda}(E) = \rho_{a,\lambda}(E)+\rho_{b,\lambda}(E)$, and hence approximating $\rho_{\lambda}(E)$ as a single Gaussian may become very inaccurate, leading to significant overestimation of the rate. An important corollary of this is that the correlation function at $\lambda_{\mathrm{sp}}$ is also poorly approximated by a Gaussian. Since the second order cumulant approximation in Eq.~(\ref{Wolynes_Correlation}) is exact at short time for any value of $\lambda$, this means that the exact correlation function, $c(t+i\lambda_{\mathrm{sp}}\hbar)$, typically exhibits significant oscillatory behaviour leading to a large reduction in the rate compared to using the short time approximation. 

Richardson and coworkers have also suggested that for a complex multidimensional reaction such as ferrous-ferric electron transfer, there are likely to be many reaction pathways and hence Wolynes theory is likely to significantly overestimate the true quantum rate.\cite{Fang20} This is a significant suggestion, because if true it would imply that the picture offered by the spin-boson model was not only quantitatively inaccurate, but also qualitatively incorrect. One of the key questions which we shall address here is therefore whether or not Wolynes theory does indeed give an inaccurate description of ferrous-ferric electron transfer.

\subsection{GR-QTST}
An alternative imaginary-time path-integral based approach, which aims to overcome the shortcomings of Wolynes theory, is the golden-rule quantum transition state theory (GR-QTST) approximation of Richardson and coworkers.\cite{Thapa19,Fang19} This is based on introducing an energy matching constraint into the path-integral simulation which is satisfied by the semi-classical instanton. The resulting approximation to $\rho_{\lambda}(0)$ can be written as
\begin{equation}
\rho_{\text{GR-QTST},\lambda}(0) =  \bigg\langle\delta\bigg(\frac{2}{3}\Big(\mathcal{E}^{(\lambda)}_0-\mathcal{E}^{(\lambda)}_1\Big)\bigg)\bigg\rangle_\lambda,
\end{equation}
where $\mathcal{E}^{(\lambda)}_i$ is a virial estimator for the energy on surface $i$ averaged around the appropriate segment of the imaginary-time path in a path-integral discretisation of the expression for $e^{-\beta F(\lambda)}$ in Eq.~(\ref{imaginary_cff}).  (Full details of the constraint functional are given in Ref.~\onlinecite{Thapa19}.) Just as with Wolynes theory, GR-QTST is expected to be most accurate when evaluated at $\lambda_{\mathrm{sp}}$, and hence the GR-QTST approximation to the rate can be written as
\begin{equation}
k_{\text{GR-QTST}} = \frac{2\pi\Delta^2}{Q_r\hbar} \rho_{\text{GR-QTST},\lambda_{\mathrm{sp}}}(0) e^{-\beta F(\lambda_{\mathrm{sp}})}, \label{GR-QTST_rate}
\end{equation}
where for ferrous-ferric electron transfer the saddle point is known by symmetry to be at $\lambda_{\mathrm{sp}}=\beta/2$. As well as being just as simple to apply to fully atomistic models as Wolynes theory, Eq.~(\ref{GR-QTST_rate}) formally reduces to Eq.~(\ref{Classical_Golden_Rule}) in the high temperature limit, and it is exact at all temperatures for a one dimensional model consisting of two linear crossing potentials.\cite{Thapa19}

Application of GR-QTST and Wolynes theory to model systems designed to exhibit multiple transition states has shown that, while Wolynes theory can overestimate the rate by as much as an order of magnitude or more, GR-QTST gives good estimates of the exact quantum rates.\cite{Fang19} In a more recent study of ferrous-ferric electron transfer, it was found that the Wolynes rate was a factor of 6 larger than the GR-QTST rate.\cite{Fang20} As we have already discussed, this was used to argue that the Wolynes theory approximation was not valid for this system, and that the true reaction was not well described by a spin-boson model.\cite{Fang20} However, as we have recently stressed,\cite{Lawrence20c} GR-QTST has a size inconsistency issue which can lead to large errors in its predictions of rates for multi-dimensional reactions. In the following we will demonstrate that this size inconsistency does in fact cause a large error in the GR-QTST rate for ferrous-ferric electron transfer, and hence that this method cannot be used to decide whether Wolynes theory is accurate for this problem.

\subsection{Linear Golden-Rule approximation}
We have recently suggested an alternative method to Wolynes theory and GR-QTST which is based on the exact result in the limit of two linear crossing diabats: the linear golden-rule (LGR) approximation.\cite{Lawrence20c}  Importantly, this does not suffer from the size inconsistency issue of GR-QTST, and it does not make the Gaussian assumption of Wolynes theory. It therefore provides an independent way to assess whether the assumptions of Wolynes theory break down in ferrous-ferric electron transfer.  The LGR approximation to the exact $\rho_\lambda(E)$ can be written as\cite{Lawrence20c} 
\begin{equation}
\rho_{\mathrm{LGR},\lambda}(E) = \big\langle\delta( \bar{V}^{(\lambda)}_{-} + \bar{\mathcal{K}}^{(\lambda)}_{-}+E)\big\rangle_\lambda,
\end{equation}
where $\bar{V}^{(\lambda)}_{-}$  is the diabatic energy gap averaged over the two bridging beads in the path-integral discretisation of $e^{-\beta F(\lambda)}$, and $\bar{\mathcal{K}}^{(\lambda)}_{-}$ is a correction term which uses the derivatives of the diabatic potentials projected along the energy gap coordinate to give a local quantum correction. Whereas the argument of the delta function in GR-QTST constrains the ring-polymer beads on each diabatic state to have the same average virial energy, LGR uses a local harmonic approximation to constrain the difference in the diabatic potential energies projected along the diabatic energy gap coordinate. Although LGR does not have the same connection to the semi-classical instanton as GR-QTST, it is expected to be more accurate in complex multidimensional systems, as the projection onto the energy gap coordinate fixes the size inconsistency of GR-QTST.\cite{Lawrence20c} 
 Again the LGR rate is evaluated at $\lambda_{\mathrm{sp}}$, where it is expected to be most accurate.  Full details of the functional forms for both  $\bar{V}^{(\lambda)}_{-}$ and $\bar{\mathcal{K}}^{(\lambda)}_{-}$  are given in Ref.~\onlinecite{Lawrence20c}.

\section{Computational Details} \label{Computational_Section}
The model of ferrous-ferric electron transfer we consider here is chosen to be the same as that considered in the study of Richardson and coworkers in which they found a large difference between the GR-QTST and Wolynes theory rates.\cite{Fang20} The system consists of 265 water molecules in a periodic cubic box of length 20~$\AA$ at 300~K. The $\text{Fe}^{2+}$ and $\text{Fe}^{3+}$ ions are fixed at a distance of 5.5~$\AA$ in the centre of the box, with the internuclear axis parallel to the edge of the box. The water model is the flexible q-TIP4P/F model (designed for use with path-integral simulations),\cite{Habershon09} and the interatomic potential for the water-Fe interaction is the same as in the original study by Chandler and coworkers.\cite{Kuharski88,Bader89,Bader90} As was discussed by Richardson and coworkers, this model is not designed to give a quantitatively accurate description of the real system. In addition to artificially fixing the positions of the Fe ions, the lack of polarisability and the relatively small size of the system are expected to lead to significant errors in the predicted reorganisation energy.\cite{Marchi93,Song93b,Blumberger08}  Despite this, the model provides a useful test of methods for calculating reaction rates in complex systems, and a qualitative representation of the physical system that will be sufficient to address the key questions of this paper (whether the assumptions of Wolynes theory are valid, and whether ferrous-ferric electron transfer can be mapped onto the spin-boson model).   

The system was first equilibrated for 1~ns in the classical canonical ensemble, from which we took 10 configurations equally spaced in time from the final 0.5~ns to initialise the path-integral simulations. Following Ref.~\citenum{Fang20} all path-integral simulations were  found to be converged using $n=24$ beads, and a time step of $0.5$~fs was used throughout. Each of the 10 initial configurations was equilibrated in the path-integral ensemble for a further 7~ps before the production runs were performed. Calculations of $F'(\lambda)$ were performed for 13 equally spaced points with $\lambda\in[0,\beta/2]$. At each value of $\lambda$ an additional 7~ps of equilibration was performed starting from the final configuration from the previous value of $\lambda$. All the necessary ensemble averages were then accumulated in a  final production run at each value of $\lambda$ amounting to 700~ps of simulation time.  All path-integral simulations were performed using the Cayley integrator with a Langevin thermostat.\cite{Korol19,Korol20}      

We find the Marcus reorganisation energy calculated using  Eq.~(\ref{Lambda_Egap}) to be $\Lambda=110.3\pm0.3$ mHartree, and that calculated using Eq.~(\ref{Lambda_def_2}) to be $\Lambda_2=113.7\pm0.3$ mHartree. Both are the same to within the error bars as those found by Richardson and coworkers.\cite{Fang20}  The close agreement between the two definitions of the reorganisation energy indicates that in the classical limit ferrous-ferric electron transfer is well described by the spin-boson model, in agreement with previous studies.\cite{Kuharski88,Bader89} As discussed above the large value of the reorganisation energy for the present model, 3.0~eV, compared to the experimental estimate of $2.1$~eV, can be attributed in large part as being due to the lack of polarisability of the q-TIP4P/F water model.\cite{Marchi93,Song93b,Blumberger08} 

\section{In defence of Wolynes theory} \label{Results_Section1}
In this section we shall address the recent finding, in the study by Richardson and coworkers,\cite{Fang20} of a large discrepancy between the rates predicted by Wolynes theory and GR-QTST for ferrous-ferric electron transfer. This was used to argue that the assumptions of Wolynes theory may be breaking down for this system, leading to an overestimate of the exact rate. Here we will demonstrate that the opposite is true. In fact it is GR-QTST which is breaking down and the assumptions of Wolynes theory are well justified. We begin by demonstrating that the lack of size consistency in GR-QTST dominates its estimate of the rate, before using a comparison of the results of Wolynes theory and LGR to demonstrate that the Wolynes theory result is highly accurate for this system.

\subsection{Size consistency error in GR-QTST}
As we have shown in a recent paper, GR-QTST is not size consistent: its prediction of the rate can be affected by degrees of freedom which are entirely uncoupled from the reaction. We found that this issue has only a relatively small effect on the GR-QTST rate for low dimensional models, however we expect it to become much more pronounced for realistic simulations which contain thousands of degrees of freedom. We thus believe that this size consistency issue is leading GR-QTST to underestimate the exact rate for the model of ferrous-ferric electron transfer described in Sec.~III.

To understand this issue we need to consider the pre-factor in the GR-QTST rate. To this end let us define the probability distribution
\begin{equation}
p_{\text{GR-QTST},\lambda}(E) =   \bigg\langle\delta\bigg(\frac{2}{3}\Big(\mathcal{E}^{(\lambda)}_0-\mathcal{E}^{(\lambda)}_1\Big)+E\bigg)\bigg\rangle_\lambda, \label{GRQTST_p_dist}
\end{equation}
for which $p_{\text{GR-QTST},\lambda}(0)=\rho_{\text{GR-QTST},\lambda}(0)$. In order to evaluate the probability distribution $p_{\text{GR-QTST},\lambda}(E)$, one simply histograms the constraint functional $\frac{2}{3}\big(\mathcal{E}^{(\lambda)}_0-\mathcal{E}^{(\lambda)}_1\big)$ in the appropriate ensemble. The lack of size consistency of GR-QTST arises because the value of $p_{\text{GR-QTST},\lambda}(0)$ -- the probability density for the virial energy on one half of the ring polymer to be the same as that on the other -- is in general dependent on all of the degrees of freedom in the simulation, and not just those which are involved in the reaction. To see why this leads to a lack of size consistency note that, even for an electronically adiabatic reaction with only one electronic state, the two halves of the ring polymer do not have the same instantaneous virial energies. Because the spontaneous energy fluctuations are size extensive, for a large enough system, the difference in virial energy between the two halves of the ring polymer will be dominated by these fluctuations rather than the difference between the diabatic potentials. Therefore, instead of constraining the system to the transition state, the GR-QTST constraint becomes dominated by quantum energy fluctuations in degrees of freedom entirely uncoupled from the reaction.

To show that this lack of size consistency is dominating the GR-QTST calculation for ferrous-ferric electron transfer, we need to demonstrate that the spontaneous fluctuations are indeed dominating $p_{\text{GR-QTST},\lambda}(E)$. To do this we consider a modified system in which the reactant and product electronic states are the same, and hence the nuclear dynamics are uncoupled from any electron transfer reaction. We define the nuclear potentials for this uncoupled system as the average of the two diabatic potentials in the real system,
\begin{equation}
V_{\mathrm{u},0}(\bm{q}) = V_{\mathrm{u},1}(\bm{q}) = \frac{V_{0}(\bm{q})+V_{1}(\bm{q})}{2},
\end{equation}
which of course just corresponds to both iron ions being $\mathrm{Fe}^{2.5+}$. Since both electronic states are the same, $\hat{H}_{\mathrm{u},0}=\hat{H}_{\mathrm{u},1}$, and the exact distribution for the uncoupled model is simply $\rho_{\lambda}(E)=\delta(E)$ [see Eq.~(\ref{Exact_distribution})]. It is clear that GR-QTST cannot satisfy this exactly, but in order for it to be accurate when applied to the real ferrous-ferric electron transfer, it must hold approximately. This means that we must have $p_{\text{GR-QTST},\lambda}(E)\simeq\delta(E)$ for the uncoupled model, i.e.~the variance must be much smaller than the variance of the exact distribution (which is on the order of  $\Lambda/\beta$).

\begin{figure}[t]
 \resizebox{1.0\columnwidth}{!} {\includegraphics{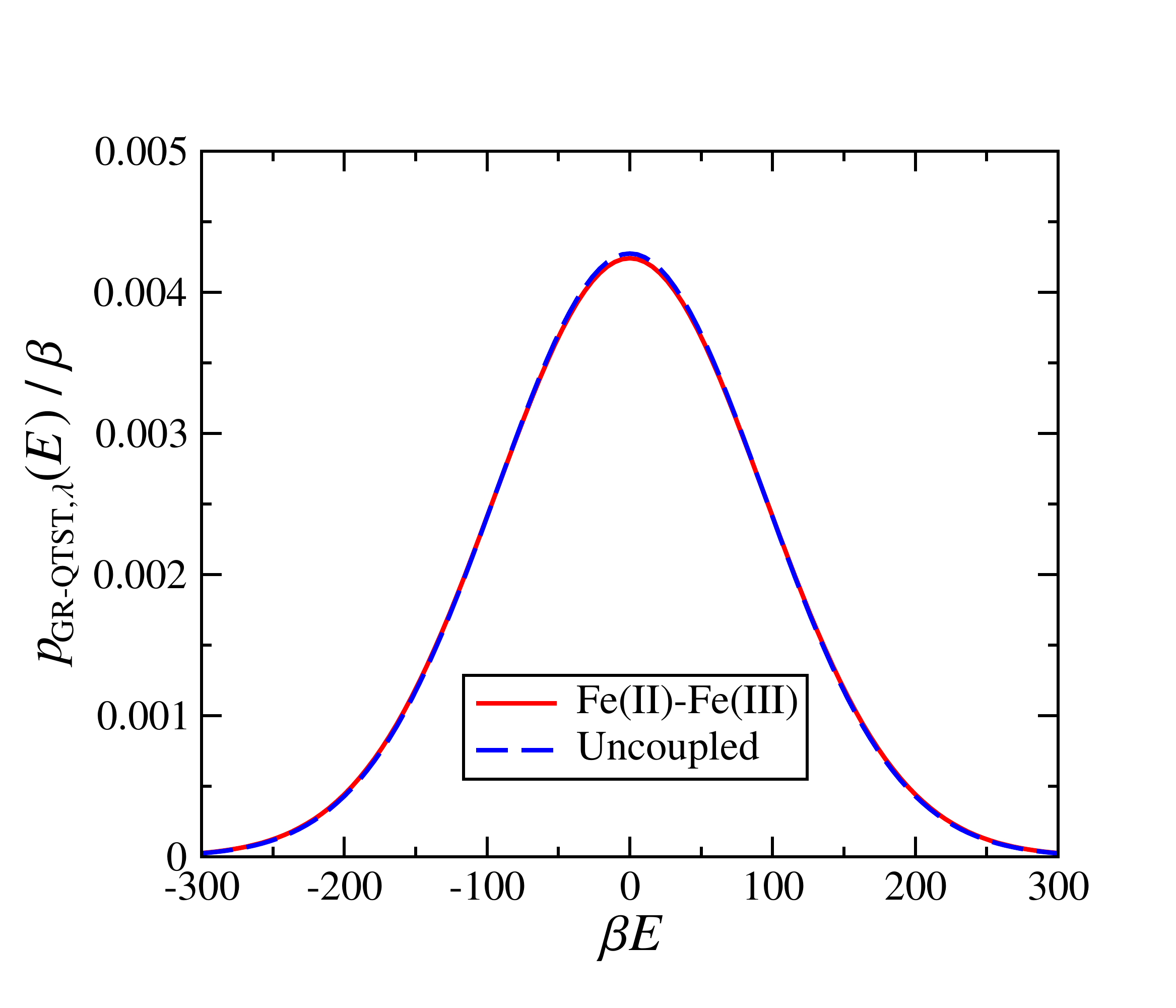}}
 \centering
 \caption{Comparison of the GR-QTST distribution (Eq.~\ref{GRQTST_p_dist}) for both the real Fe(II)-Fe(III) system and for an uncoupled system in which the nuclear potentials are the same in both diabatic states (corresponding to both iron ions having a charge of 2.5). The close agreement between the two curves illustrates that the size consistency error of GR-QTST is dominating the calculation of the rate in ferrous-ferric electron transfer. Note for reference that $\beta\Lambda\simeq116$.}
 \label{GRQTST_Dist}
 \end{figure}

Figure \ref{GRQTST_Dist} compares the distribution $p_{\text{GR-QTST},\lambda}(E)$ for both the real ferrous-ferric system and for the uncoupled system, in which the two diabatic electronic states are the same. We see that, not only is the variance of $p_{\text{GR-QTST},\lambda}(E)$ in the uncoupled system not significantly smaller than in the real system, but it is actually essentially the same, indicating that spontaneous quantum fluctuations in the energy are dominating the GR-QTST constraint functional and hence the GR-QTST approximation of $\rho_{\lambda}(0)$. This demonstrates clearly that the GR-QTST rate is not a reliable approximation to the exact rate for this system, and hence cannot be used to assess the accuracy of Wolynes theory or mapping to the spin-boson model.

\subsection{Accuracy of Wolynes theory}
Although we have demonstrated that the GR-QTST rate is not reliable for ferrous-ferric electron transfer this does not, by itself, disprove the suggestion that Wolynes theory is overestimating the exact rate. One way to address this possibility is to apply a method which does not assume that the correlation function is Gaussian and compare the results with Wolynes theory. While GR-QTST cannot be used for this purpose, we can apply the LGR approximation, which although closely related to GR-QTST is size consistent. In addition to providing a useful way to independently assess the assumptions of Wolynes theory, this will also serve as the first demonstration of the applicability of LGR to a fully atomistic simulation.

Table \ref{rates_table}  compares the rate constants for ferrous-ferric electron transfer calculated using each of the methods under consideration. Just as in the study by Richardson and coworkers,\cite{Fang20} we find that Wolynes theory predicts a rate which is around 7 times larger than that predicted by GR-QTST. In contrast, the Wolynes rate is only about 5\% larger than the LGR rate. The close agreement of Wolynes theory and LGR, which are based on very different approximations, is a strong indication that the assumptions of Wolynes theory are likely to be valid for ferrous-ferric electron transfer. We would also note that the errors given in Table \ref{rates_table} are dominated by the thermodynamic integration of $F'(\lambda)$ to give $e^{-\beta F(\lambda_{\mathrm{sp}})}$, and hence we can give a much more detailed comparison of the methods by separately considering their approximations to $\rho_{\lambda_{\mathrm{sp}}}(E)$.

\begin{table}[t] 
\renewcommand\arraystretch{1.5}
\caption{Comparison of the rate constants for each of the methods considered, given in atomic units. Note that errors are given as two standard errors in the mean.}
\vspace{0.25cm}
\begin{tabular}{ccccc}
\toprule
  & $ k / \Delta^2$                                           \\ 
\hline\hline
Classical & $5.4\pm0.6\times 10^{-11}$                                   \\
Marcus  & $4.3\pm0.4\times 10^{-11}$                           \\
Wolynes & $ 3.9\pm0.6\times 10^{-9}$                                 \\
LGR & $3.7\pm0.6\times 10^{-9}$                                    \\
SB Mapping & $2.8\pm0.3\times 10^{-9}$                                   \\
GR-QTST & $5.3\pm0.8\times 10^{-10}$                             \\
\hline
\end{tabular} \label{rates_table}
\end{table}

\begin{figure}[t]
 \resizebox{1.0\columnwidth}{!} {\includegraphics{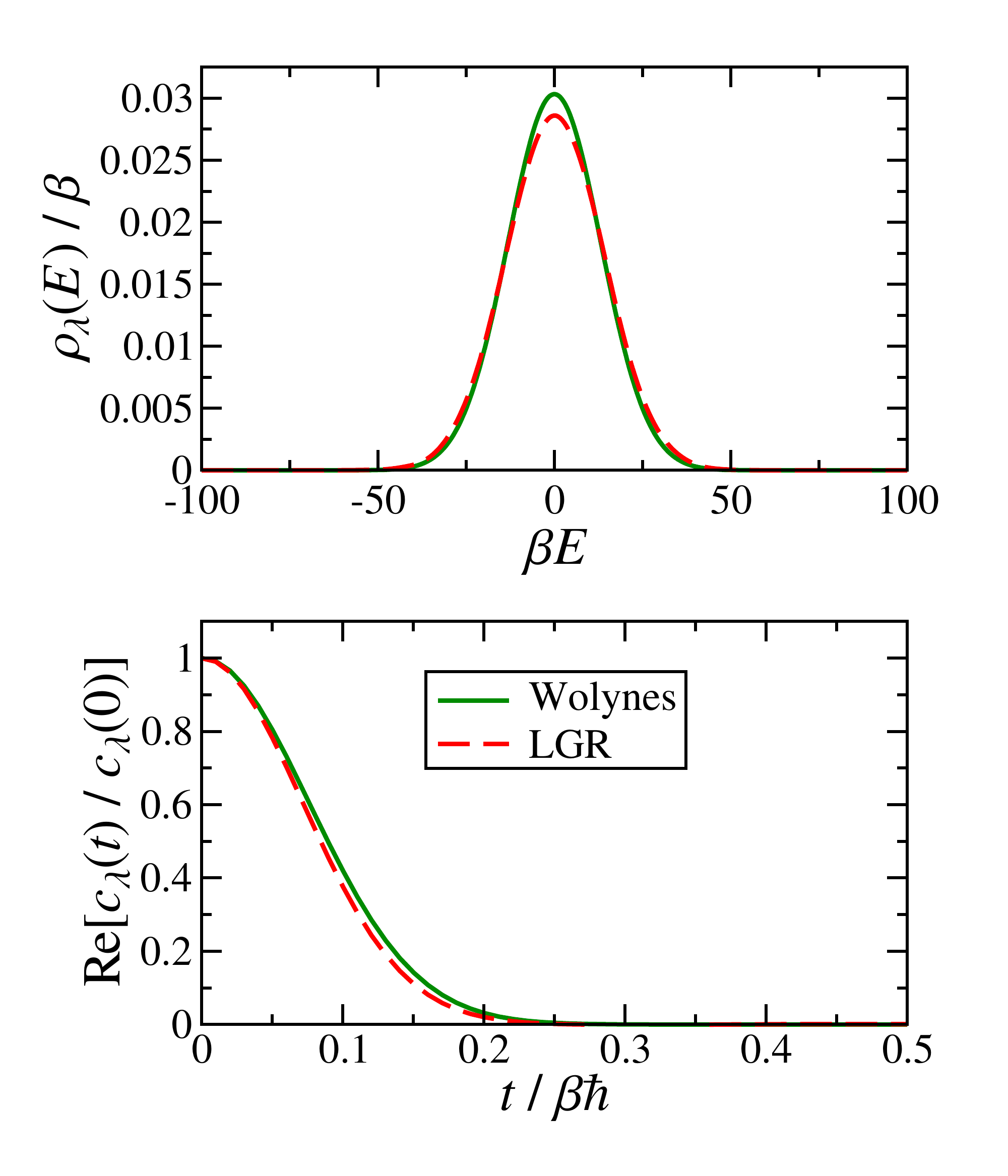}}
 \centering
 \caption{LGR and Wolynes theory approximations to $\rho_{\lambda}(E)$ (upper panel) and 
 $\Re[c_\lambda(t)/c_\lambda(0)]$ (lower panel) for the present atomistic model of ferrous-ferric electron transfer. In both panels calculations are done with $\lambda=\lambda_{\mathrm{sp}}=\beta/2$, using 24 path-integral beads. }
 \label{LGR_Woly_Dist_Correl}
 \end{figure}

The upper panel of Fig.~\ref{LGR_Woly_Dist_Correl} compares the LGR and Wolynes theory approximations to $\rho_\lambda(E)$ with $\lambda=\lambda_{\rm sp}=\beta/2$. It is clear that both agree very closely, with LGR having an approximately Gaussian functional form. Using the fact that the distribution is just the Fourier transform of the correlation function we can also compare the Wolynes and LGR approximations to the normalised correlation function, by taking the inverse Fourier transform to give
\begin{equation}
\frac{c_{\lambda}(t)}{c_{\lambda}(0)} = \int_{-\infty}^{\infty} \rho_{\lambda}(E) e^{+iE t/\hbar} \mathrm{d}E. 
\end{equation}
The LGR and Wolynes correlation functions are shown in the lower panel of Fig.~\ref{LGR_Woly_Dist_Correl} and unsurprisingly we find that the LGR correlation function looks essentially like a Gaussian in the time domain as well, again indicating that the assumptions of Wolynes theory are valid. The small 5\% difference in the rates of the two methods is visible in both panels, from the relative heights of the distributions and areas under the correlation functions. What is difficult to see by eye is whether this difference is due to the LGR correlation function not being perfectly Gaussian, or if it is due to a small error in the LGR approximation to the second moment of the exact distribution in Eq.~(\ref{Second_moment}). 

Since LGR does not in general give the exact first and second moments of the distribution $\rho_{\lambda}(E)$,\cite{Footnote3} one test of its accuracy is to compare the rates obtained using the original LGR distribution with those obtained using a corrected LGR distribution in which the first and second moments are adjusted to the exact results. This can be achieved by defining the ``moment corrected LGR'' distribution according to
\begin{equation}
\rho_{\text{mc-LGR},\lambda}(E) =\sqrt{\frac{\tilde{\mu}_{2,\lambda}}{\mu_{2,\lambda}}}\rho_{\mathrm{LGR},\lambda}\Bigg(\sqrt{\frac{\tilde{\mu}_{2,\lambda}}{\mu_{2,\lambda}}}(E-\mu_{1,\lambda})+\tilde{\mu}_{1,\lambda} \Bigg)
\end{equation}
where
\begin{equation}
\tilde{\mu}_{1,\lambda} = \int_{-\infty}^{\infty} E \rho_{\mathrm{LGR},\lambda}(E) \mathrm{d}E 
\end{equation}
and
\begin{equation}
\tilde{\mu}_{2,\lambda} = \int_{-\infty}^{\infty} (E^2-\tilde{\mu}_{1,\lambda}^2) \rho_{\mathrm{LGR},\lambda}(E) \mathrm{d}E .
\end{equation}
We find that the moment corrected LGR rate agrees with the Wolynes theory rate to within 1\%, indicating that the small difference between the LGR and Wolynes theory rates is almost entirely due to the small error in the LGR approximation to $\mu_{2,\lambda}$. This therefore strongly supports the assertion that the Wolynes theory approximation is valid for ferrous-ferric electron transfer.

It is now clear that the difference between the GR-QTST rate and the Wolynes rate for ferrous-ferric electron transfer is due to GR-QTST underestimating the exact rate rather than Wolynes theory overestimating it. The error in GR-QTST originates from the fluctuations in the virial energies of modes uncoupled to the reaction, which dominate the GR-QTST constraint functional. Given this perspective it seems natural to ask whether these background fluctuations can be subtracted from the GR-QTST calculation to yield an accurate rate. In particular, the above discussion would seem to suggest that the small difference between the GR-QTST distribution for the real system and the uncoupled system in Fig.~\ref{GRQTST_Dist} should be related to the ``correct'' GR-QTST result. That is, the result that would be obtained if the contribution from the uncoupled modes could be removed. A detailed discussion of how this can be done is left to Appendix \ref{Fixing_GR-QTST}, but the result is that one can indeed approximately remove the spurious background signal from the GR-QTST distribution. In doing so, we find that the resulting corrected GR-QTST rate is about $10\%$ larger than the Wolynes rate, rather than 7 times smaller. This further confirms our assertion that it is GR-QTST and not Wolynes theory that is breaking down for this problem.

\section{Validity of mapping to the spin-boson model} \label{Results_Section2}
The study by Richardson and coworkers also concluded that the common practice of mapping to the spin-boson model is invalid for ferrous-ferric electron transfer.\cite{Fang20} In this section we give an overview of their arguments, and discuss how their results can be reinterpreted in the light of our present findings. We then go on to reassess the validity of mapping to the spin-boson model for ferrous-ferric electron transfer and discuss what conclusions can be drawn for more general systems. When it comes to assessing the validity of the mapping there are two main questions one might ask: is the spin-boson perspective qualitatively accurate, and can it give quantitatively accurate predictions of the rate?

Richardson and coworkers used the difference between the Wolynes theory results and GR-QTST results to argue that mapping to the spin-boson model was invalid.\cite{Fang20} Their argument focussed on the close agreement between GR-QTST and Wolynes theory in a previous study, which consisted of a series of different spin-boson models, as evidence that the ferrous-ferric system could not be spin-boson like. However, that previous study only looked at spin-boson models with up to 8 degrees of freedom,\cite{Thapa19} compared to the over 2000 degrees of freedom in their atomistic model of ferrous-ferric electron transfer.\cite{Fang20} As we have demonstrated in Ref.~\citenum{Lawrence20c}, the size consistency error of GR-QTST only becomes apparent when there a large number of uncoupled degrees of freedom. Hence the difference between GR-QTST and Wolynes theory for ferrous-ferric electron transfer cannot be used to argue that mapping to a spin-boson model is invalid for this problem.

\begin{table}[t] 
\renewcommand\arraystretch{1.5}
\caption{Quantum correction factors for the different methods considered in this paper. In order to provide a fair comparison, the quantum correction factors for the atomistic methods are defined relative to the atomistic classical golden-rule rate, whereas the spin-boson quantum correction factor is defined relative to the classical golden-rule rate for the same spin-boson model (i.e. Marcus theory). }
\vspace{0.25cm}
\begin{tabular}{ccccc} 
\toprule
  & $\Gamma$                                            \\ 
\hline\hline
Wolynes &    $73\pm14$                                  \\
LGR &  $69\pm 13$                                    \\
SB Mapping &   $66\pm2$                                   \\
GR-QTST &   $10\pm2$                                \\
\hline
\end{tabular} \label{Quantum_Correc_table}
\end{table}

In order to provide further evidence that Wolynes theory was breaking down in their ferrous-ferric simulation, and to justify the claim that it is not valid to map this problem onto a spin-boson model, Richardson and coworkers also performed an instanton analysis.\cite{Fang20} They took a series of configurations from an equilibrium simulation and fixed all atoms more than $5\AA$ away from the Fe ions, before performing an instanton optimisation on the remaining degrees of freedom. They found that the resulting instantons had a wide range of different saddle point values, $\lambda_{\mathrm{sp}}$, which they used to argue that the system exhibits a continuum of different transition states with different tunnelling characters.\cite{Fang20} However, we note that if one were to do the same thing to the spin-boson model, for example by fixing the coordinates of some of the bath modes at configurations sampled from an equilibrium simulation, one would also see a range of different saddle points. Fixing some degrees of freedom stops them from relaxing and reduces the reorganisation energy, while also giving a random bias either to products or reactants. The different biases then lead to a range of different saddle points, and the distribution of these saddle points is further broadened by the reduced reorganisation energy. Since the spin-boson model is known to have just a single instanton at any given temperature, this seems to us to refute the claim that the same analysis can be used to show that the ferrous-ferric problem has multiple transition states.

With this established, let us now return to the qualitative Marcus picture of electron transfer, in which collective solvent motion takes the system to the transition state. This picture is strongly supported by the close agreement between the classical golden-rule and Marcus theory rates, and the Wolynes theory and quantum mechanical spin-boson rates, in Table \ref{rates_table}. However, this close agreement is not perfect in either (classical or quantum) case: we find small but significant quantitative differences between the atomistic and spin-boson-mapped calculations, reflecting the fact that the real potential energy surface is anharmonic. 

Anharmonic effects are also apparent in the difference between the two definitions of the Marcus reorganisation energy, $\Lambda$ and $\Lambda_2$ [Eqs.~(\ref{Lambda_Egap}) and (\ref{Lambda_def_2}), respectively]. Although there is only a $3\%$ difference between the two, the rates predicted by $\Lambda_2$ are significantly less accurate -- more than a factor of 2 smaller in both the classical and quantum cases -- because using $\Lambda_2$ overestimates the activation energy of the reaction. This highlights the fact that the quality of the mapping is strongly dependent on how well it captures the activation energy. In fact, if we adjust the reorganisation energy so that Marcus theory agrees with the exact classical golden-rule rate, this also brings the quantum rate predicted by the spin-boson model up to agree with the Wolynes and LGR predictions to within their error bars, giving $k/\Delta^2  \simeq 3.5\times10^{-9}$ atomic units. Hence, we find that accounting for the small anharmonic effects in an average manner can make mapping to the spin-boson model essentially quantitative for this model of ferrous-ferric electron transfer.

Table II shows the quantum correction factors, $\Gamma=k/k_{\mathrm{cl}}$, obtained using each of the methods discussed in this paper. It is worth noting that the quantum correction factors predicted by Wolynes theory, LGR and the spin-boson mapping, are much larger than the experimental estimates. As we have now demonstrated the accuracy of these approaches, it follows that this difference is caused by deficiencies in the atomistic model we have used, and as has been discussed at length before this is likely due in large part to a lack of polarisability in the water model.\cite{Marchi93,Song93b,Blumberger08} The inclusion of polarisability is expected to significantly reduce the reorganisation energy and hence the importance of tunnelling in the system. These changes are likely to make the assumptions behind mapping to a spin-boson model more rather than less reliable, and hence we expect that the results found here are transferable to more realistic models.


Despite the fact that we find mapping to the spin-boson model can be quantitatively accurate for ferrous-ferric electron transfer, our results have highlighted the sensitivity of the mapping to the precise approach used. It is clear that for any system in which there are modes involved in the reaction that are strongly anharmonic the mapping will be less reliable. Furthermore, even for systems in which the reactants and products can be treated as essentially harmonic, there may be situations in which they have different frequencies, which will again make the spin-boson model less reliable. While there exist many asymmetric generalisations of basic Marcus theory,\cite{Marcus65,Laborda13,Mattiat18} the importance of nuclear quantum effects in electron transfer can be very pronounced, and hence for any theory to be quantitatively accurate it must be able to capture both asymmetry and nuclear tunnelling. Unlike mapping to the spin-boson model the path-integral methods we have used in this paper have no difficulty in dealing with more complicated systems, and hence provide a useful tool in the study of complex electron transfer reactions in solution. 
  
\section{Conclusion} \label{Conclusion}
For at least 30 years the standard picture of ferrous-ferric electron transfer has been based on linear response theory. While this has been rigorously tested in the high temperature limit where it is possible to calculate all quantities exactly,\cite{Kuharski88,Bader89,Bader90} it has previously just been assumed that the same approach would carry over to the quantum regime. This assumption has recently been thrown into doubt by the study of Richardson and coworkers,\cite{Fang20} which has suggested that when nuclear tunnelling is important the assumptions of linear response theory may break down due to the existence of a range of tunnelling paths with qualitatively different behaviour. This would imply that, not only was the common practice of mapping electron transfer reactions onto the spin-boson model quantitatively wrong, but also qualitatively so. In this paper we have addressed these suggestions and confirmed that the traditional  assumption that ferrous-ferric electron transfer is well described by linear response is well justified, even in the presence of significant tunnelling. 

Richardson and coworkers have previously shown that the saddle point approximation of Wolynes theory can breakdown for systems which exhibit multiple distinct transition states.\cite{Fang19} By showing that GR-QTST and Wolynes theory give very different predictions of the electron transfer rate in ferrous-ferric electron transfer, they argued that this may be due to the existence of multiple transition states, which would be at odds with the assumptions of linear response and hence mapping onto the spin-boson model. Here we have demonstrated that the difference between GR-GTST and Wolynes theory is due to a size consistency error on the part of GR-QTST, and by comparing with an alternative path-integral method, LGR, we have confirmed the accuracy of Wolynes theory. Just as in the original study by Chandler and coworkers,\cite{Bader90} we have found a tunnelling enhancement that is much larger than inferred from experiment. This is most likely due to the lack of polarisability in the water models used in this study and in Ref.~\citenum{Bader90}, and also to the relatively small sizes of the simulations. The model we have investigated is thus not entirely realistic. However, it is clear from the error it has revealed in GR-QTST that it provides a useful test for methods designed to treat non-adiabatic systems, and given that the Wolynes theory results are essentially exact for this model it could provide a useful benchmark for testing other methods in the future.  

Comparison of our path-integral results with those from mapping to the spin-boson model show small deviations from the harmonic assumption. We find that these deviations are sufficiently small that they can be essentially accounted for, in an average sense, by choosing the Marcus reorganisation energy so as to give the correct classical activation energy. However our results highlight that even small anharmonic effects can lead to significant changes in the rate, and we note that these effects are likely to be much more pronounced in systems which are more complex than aqueous ferrous-ferric electron transfer. In particular if the reactants and products have different frequencies or if there are modes important in the electron transfer which show more pronounced anharmonicity, such as conformational changes which couple to the electron transfer, then mapping onto the spin-boson model is likely to breakdown.\cite{Blumberger06b,Krapf12} In contrast path-integral approaches such as Wolynes theory and the LGR approximation provide a simple approach to the calculation of electron transfer rates in complex systems.

\acknowledgments
We would like to thank Jeremy Richardson for his comments on this manuscript. J. E. Lawrence is supported by The Queen's College Cyril and Phillis Long Scholarship in conjunction with the Clarendon Fund of the University of Oxford and by the EPRSC Centre for Doctoral Training in Theory and Modelling in the Chemical Sciences, EPSRC grant no. EP/L015722/1.

\section*{Data Availability Statement}
The data that support the findings of this study are available in the paper itself. 

\appendix
\section{Mapping to the spin-boson model} \label{Mapping_to_SB_apdx}
In this appendix we give a discussion of the basis for mapping electron transfer onto a spin-boson model via classical equilibrium simulation. In order to derive the standard mapping we begin by writing the rate in the form
\begin{equation}
k = \frac{\Delta^2}{\hbar^2} \int_{-\infty}^{\infty} \Big\langle e^{-i\hat{H}_0 t/\hbar}e^{+i\hat{H}_1 t/\hbar} \Big\rangle_0 \mathrm{d}t,
\end{equation}
where $\langle \hat{A} \rangle_0=\tr[e^{-\beta\hat{H}_0}\hat{A}]/Q_r$. Noting that
\begin{equation}
e^{-i\hat{H}_0 t/\hbar}e^{+i\hat{H}_1 t/\hbar} = \mathcal{T}\!\!\exp(-\frac{i}{\hbar}\int_0^{t} e^{-i\hat{H}_0 t'/\hbar}\hat{V}_-e^{+i\hat{H}_0 t'/\hbar} \mathrm{d}t')
\end{equation}
where $\mathcal{T}\!\!\exp$ denotes the time ordered exponential, we can then make use of Kubo's generalised cumulant expansion to write
\begin{equation}
k = \frac{\Delta^2}{\hbar^2}\int_{-\infty}^{\infty} e^{-i\langle \hat{V}_{-}\rangle_0 t/\hbar  - \frac{1}{\hbar^2} \int_0^{t} (t-t')\langle \delta\hat{V}_{-}(0) \delta\hat{V}_{-}(t)\rangle_0\mathrm{d}t' + \dots} \mathrm{d}t
\end{equation}
where
\begin{equation}
\delta\hat{V}_{-}(t) = e^{+i\hat{H}_0t/\hbar}(\hat{V}_{-}-\langle \hat{V}_{-}\rangle_0)e^{-i\hat{H}_0t/\hbar}.
\end{equation}
In general there exist infinitely many non-zero cumulants in the exponential, however for the spin-boson model the series truncates at the second cumulant. Hence in order to map onto the spin-boson model one truncates at second order to give 
\begin{equation}
\frac{c(t)}{Q_r} \simeq e^{-i\langle \hat{V}_{-}\rangle_0t/\hbar  - \frac{1}{\hbar^2} \int_0^{t} (t-t')\langle \delta\hat{V}_{-}(0) \delta\hat{V}_{-}(t)\rangle_0\mathrm{d}t' }. \label{SB_map_1}
\end{equation}
To simplify this into a form which can be evaluated using classical molecular dynamics, one begins by introducing the Kubo transformed energy gap fluctuation operator
\begin{equation}
\delta\hat{V}_{-}^{(K)}  = \frac{1}{\beta}\int_0^\beta \mathrm{d}\lambda\, e^{\lambda \hat{H}_0}\delta\hat{V}_{-}e^{-\lambda \hat{H}_0}, 
\end{equation}
from which one can define the spectral density, $J(\omega)$, as
\begin{equation}
\frac{J(\omega)}{\omega}=\frac{\beta}{4}\int_0^{\infty} \big\langle \delta\hat{V}_{-}^{(K)}(0) \, \delta\hat{V}_{-}(t)\big\rangle_0  \cos(\omega t)\, \mathrm{d}t.
\end{equation}
Then making use of the relation between the standard correlation function $C(t)=\left<\delta\hat{V}_-(0)\delta\hat{V}_-(t)\right>_0$ and its Kubo transformed variant $C^{(K)}(t)=\left<\delta\hat{V}_-^{(K)}(0)\delta\hat{V}_-(t)\right>_0$,
\begin{equation}
C(t) = \frac{1}{2\pi} \int_{-\infty}^{\infty}\int_{-\infty}^{\infty}e^{i\omega (t'-t)} \frac{\beta\hbar\omega}{e^{\beta\hbar\omega}-1} C^{(K)}(t') \,\mathrm{d}t' \mathrm{d}\omega,
\end{equation} 
we can write 
\begin{equation}
\langle \delta\hat{V}_{-}(0) \delta\hat{V}_{-}(t)\rangle_0 = \frac{4}{\pi} \int_{-\infty}^{\infty} e^{-i\omega t} \frac{\hbar}{e^{\beta\hbar\omega}-1}  J(\omega)\, \mathrm{d}\omega.
\end{equation}
Finally using the relations
\begin{equation}
\int_0^{t} (t-t') e^{-i\omega t'} \mathrm{d}t' =  - \frac{i t}{\omega} + \frac{1 - e^{-i\omega t}}{\omega^2} 
\end{equation}
and
\begin{equation}
\frac{2}{e^{\beta\hbar\omega}-1} = \coth(\beta\hbar\omega/2)-1
\end{equation}
along with the definition of the reorganisation energy
\begin{equation}
\frac{4}{\pi}\int_0^{\infty} \frac{J(\omega)}{\omega} \,\mathrm{d}\omega = \Lambda, \label{Lambda_J_A}
\end{equation}
we can rewrite Eq.~(\ref{SB_map_1}) in the form
\begin{equation}
\frac{c(t)}{Q_r} \simeq \exp(-i \epsilon t/\hbar - \phi(t)/\hbar),
\end{equation}
where
\begin{equation}
\phi(t) = \frac{4}{\pi}\int_{-\infty}^{\infty}\frac{J(\omega)}{\omega^2}\Bigg[\frac{1-\cos(\omega t)}{\tanh(\beta\hbar\omega/2)}-i \sin(\omega t)\Bigg]\,\mathrm{d}\omega
\end{equation}
and the thermodynamic driving force is 
\begin{equation}
\epsilon  = \langle \hat{V}_{-}\rangle_0 + \Lambda. \label{Eff_driving_force1}
\end{equation}
 For the spin-boson model, this is the same as the free energy difference between the reactants and products in the system of interest,
\begin{equation}
\epsilon = \frac{1}{\beta}\ln(\frac{\tr[e^{-\beta\hat{H}_1}]}{\tr[e^{-\beta\hat{H}_0}]}). \label{Eff_driving_force2}
\end{equation}
 Unfortunately, as we shall discuss in more detail shortly, a general system will not satisfy Eqs.~(\ref{Lambda_J_A}), (\ref{Eff_driving_force1}) and (\ref{Eff_driving_force2}) simultaneously. However, it is clear that for an accurate mapping it is sensible to require that Eq.~(\ref{Eff_driving_force2}) is satisfied.

At this stage the only approximation that has been made is to truncate the cumulant expansion to second order in $\delta\hat{V}_{-}$, which is exact for the spin-boson model. Typically one makes a further simplification by using classical quantities in place of their quantum counterparts, which again is exact in the case of the spin-boson model. In particular, this means using the classical correlation function in place of the Kubo transformed correlation function in the definition of the spectral density
\begin{equation}
\frac{J(\omega)}{\omega}\simeq\frac{\beta}{4}\int_0^{\infty} \big\langle \delta V_{-}(0) \, \delta V_{-}(t)\big\rangle_{\mathrm{cl},0}  \cos(\omega t)\, \mathrm{d}t,
\end{equation}
which is straightforward to compute using standard MD codes. 

\begin{figure}[t]
 \resizebox{1.0\columnwidth}{!} {\includegraphics{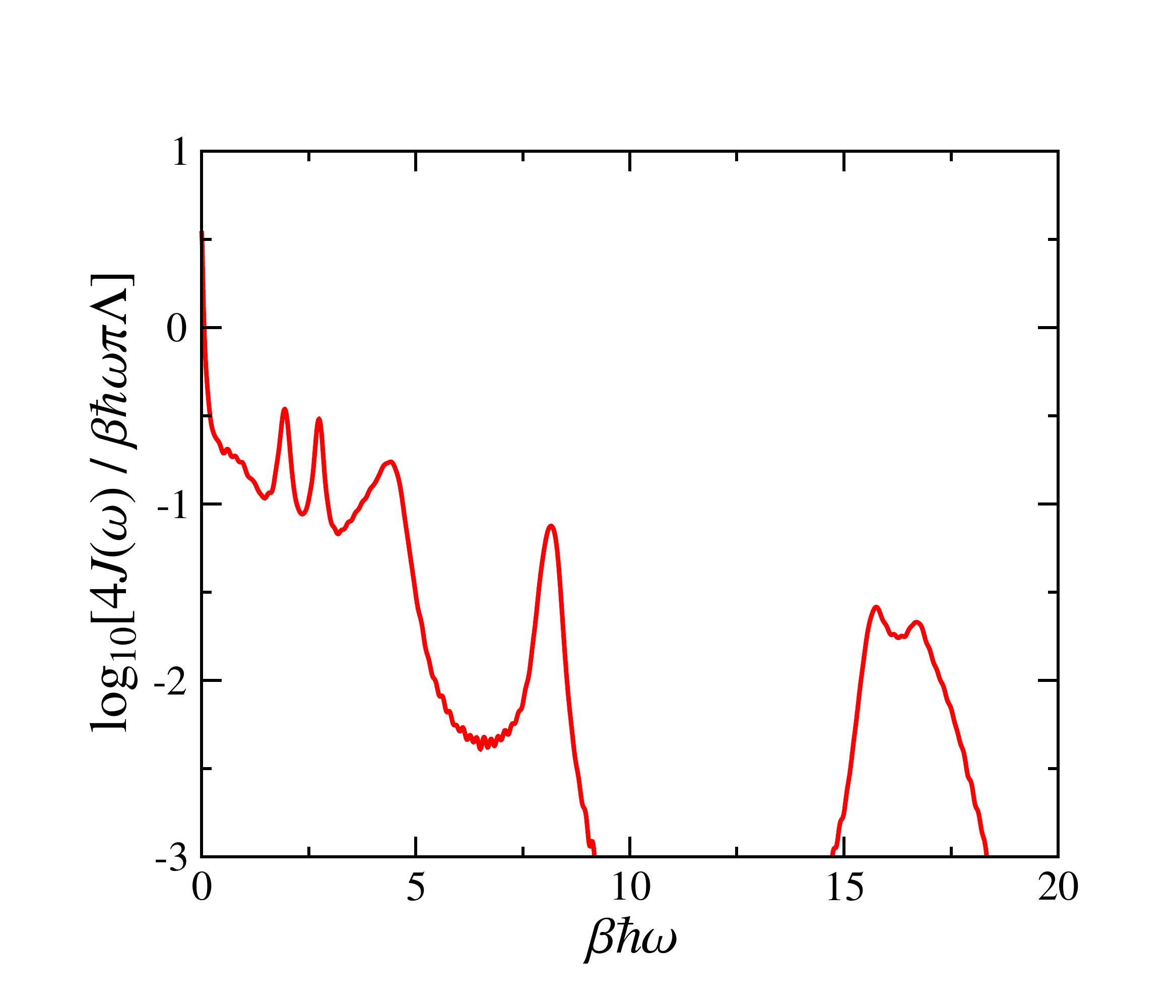}}
 \centering
 \caption{Logarithmic plot of the spectral density calculated using Eq.~(\ref{Appendix_Spec_Dens}). This was used to evaluate the spin-boson mapping results in the text.}
 \label{Classical_Spec_Density}
 \end{figure}

 Now comparing Eq.~(\ref{Lambda_J_A}) with Eqs.~(\ref{Eff_driving_force1}) and (\ref{Eff_driving_force2}), we note that there are effectively two different definitions of the Marcus reorganisation energy, which can be computed either as the difference between the driving force and mean diabatic energy gap
  \begin{equation}
 \Lambda = \epsilon - \langle {V}_{-}\rangle_{\mathrm{cl},0}, \label{Lambda_Egap_A},
 \end{equation}
 or from the integral of the spectral density
 \begin{equation}
 \Lambda_2 = \frac{\beta}{2}  \big\langle \delta V_{-}(0) \, \delta V_{-}(0)\big\rangle_{\mathrm{cl},0} .
 \end{equation}
Since the reorganisation energy appears in the exponent in Marcus theory, even small changes in it can result in large changes to the rate. This means that choosing the correct definition is important. Since mapping to the spin-boson model is equivalent to assuming that the free energy $F_{\mathrm{cl}}(\lambda)$ is parabolic, the most consistent way to define $\Lambda$ for ferrous-ferric electron transfer is such that it passes through the three known points at $\lambda=0$, $\beta/2$ and $\beta$, which corresponds to using Eq.~(\ref{Lambda_Egap_A}). Hence for consistency in the main text we define the spectral density such that
\begin{equation}
\frac{J(\omega)}{\omega}=\frac{\Lambda}{2}\int_0^{\,\infty} \frac{\big\langle \delta V_{-}(0) \, \delta V_{-}(t)\big\rangle_{\mathrm{cl},0}}{\big\langle  \delta V_{-}^2(0)\big\rangle_{\mathrm{cl},0}}  \cos(\omega t) \mathrm{d}t, \label{Appendix_Spec_Dens}
\end{equation}
which ensures that the integral of the spectral density now gives the more accurate version of the Marcus reorganisation energy in Eq.~(\ref{Lambda_Egap_A}). The spectral density, calculated using Eq.~(\ref{Appendix_Spec_Dens}), is shown in Fig.~\ref{Classical_Spec_Density}.
 
We note that alternatively one could use RPMD to approximate the Kubo transformed energy gap correlation function. While this may improve the accuracy of the calculated rates near activationless, $\epsilon\sim\Lambda$, it still does not allow one to overcome the fundamental limitations of mapping onto a spin-boson model. In particular it is fundamentally limited to sampling equilibrium configurations of the reactants, and hence cannot capture anharmonic features of the potential away from equilibrium that may be important in reaching the transition state. We also note that the above argument can equally be applied to give expressions evaluated in the product rather than the reactant ensemble. While this is the same for the spin-boson model for any realistic asymmetric system there may be significant differences in the reactant and product potential energy surfaces and this can lead to a serious failure of the mapping approach for significantly activated reactions. 

\section{Moments of rate distribution} \label{Moments_Appendix}
Here we give a brief derivation of the relation between the moments of the exact distribution $\rho_\lambda(E)$ and the derivatives of the free energy $F(\lambda)$. We begin by considering the 1st central moment of the distribution which is defined as,
\begin{equation}
\mu_{1,\lambda} = \int_{-\infty}^{\infty} E  \rho_{\lambda}(E) \,\mathrm{d}E.  
\end{equation}
Inserting the defintion of the distribution this is
\begin{equation}
\mu_{1,\lambda}   = \frac{1}{2\pi\hbar}\int_{-\infty}^{\infty}  \int_{-\infty}^{\infty}  \Big\langle e^{-i\hat{H}_0 t/\hbar}e^{+i \hat{H}_1 t/\hbar} \Big\rangle_\lambda E e^{-iE t/\hbar}\, \mathrm{d}E \mathrm{d}t.
\end{equation}
Replacing the multiplication by $E$ with a derivative with respect to time and integrating by parts leads to
\begin{equation}
\mu_{1,\lambda}   = \frac{-i}{2\pi}\int_{-\infty}^{\infty}  \int_{-\infty}^{\infty}   \Bigg\langle  \frac{\partial}{\partial t}e^{-i\hat{H}_0 t/\hbar}e^{+i \hat{H}_1 t/\hbar} \Bigg\rangle_\lambda  e^{-iEt/\hbar}\, \mathrm{d}E \mathrm{d}t,
\end{equation}
which can then be integrated over $E$ to give
\begin{equation}
\mu_{1,\lambda}   = -i\hbar  \int_{-\infty}^{\infty}   \Bigg\langle  \frac{\partial}{\partial t}e^{-i\hat{H}_0 t/\hbar}e^{+i \hat{H}_1 t/\hbar} \Bigg\rangle_\lambda  \delta(t) \,\mathrm{d}t.
\end{equation}
This then immediately simplifies to
\begin{equation}
\mu_{1,\lambda}   = -\Big\langle \hat{V}_0-\hat{V}_1 \Big\rangle_\lambda = \beta F'(\lambda), 
\end{equation}
and hence we see that the first moment is directly proportional to the first derivative of the free energy at $\lambda$.

We can apply the same procedure to evaluate the second central moment,
\begin{equation}
\mu_{2,\lambda} = \int_{-\infty}^{\infty} (E^2-\mu_{1,\lambda}^2)  \rho_{\lambda}(E)\, \mathrm{d}E,  
\end{equation}
from which it follows that
\begin{equation}
\mu_{2,\lambda} = -\hbar^2  \int_{-\infty}^{\infty}   \Bigg\langle  \frac{\partial^2}{\partial t^2}e^{-i\hat{H}_0 t/\hbar}e^{+i \hat{H}_1 t/\hbar} \Bigg\rangle_\lambda  \delta(t)\, \mathrm{d}t-\mu_{1,\lambda}^2.
\end{equation}
This then simplifies to give the second central moment in terms of the second derivative of the free energy,
\begin{equation}
\mu_{2,\lambda} =  \langle  V_{-}(\boldsymbol{q}_0)V_{-}(\boldsymbol{q}_l)\rangle_\lambda -\mu_{1,\lambda}^2 = -\beta F''(\lambda).
\end{equation}

\section{Removing the uncoupled modes from GR-QTST} \label{Fixing_GR-QTST}
Here we discuss how the spontaneous quantum fluctuations can be approximately removed from the GR-QTST approximation to the ferrous-ferric electron transfer rate. As we have discussed in a recent paper,\cite{Lawrence20c} the GR-QTST approximation to $\rho_\lambda(E)$ is affected by degrees of freedom which are not coupled to the electronic transition. As we have argued in the main text, in a large system such as the atomistic model of ferrous-ferric electron transfer there will be many such degrees of freedom, which will contaminate the GR-QTST distribution leading it to underestimate the true rate. 

Let us therefore suppose that we can identify a relatively small subset of degrees of freedom, labelled $a$, which are involved in the reaction, and a large remainder which are effectively uncoupled from the reaction, labelled $b$, such that we can write the diabatic potentials in the forms
\begin{subequations}
\begin{align}  
V_0(\bm{q}) &= U_{0,a}(\bm{q}_a) + U_b(\bm{q}_{b})  \\
V_1(\bm{q}) &= U_{1,a}(\bm{q}_a) + U_b(\bm{q}_{b}).
\end{align}
\label{Uncoupled_model}
\end{subequations}
Note that we will not actually need to find the coordinate transformation or identify the coordinates $a$ and $b$ for which this is true, but only assume that it is in principle possible.

 Having made this assumption we can simply decompose the contributions to $p_{\text{GR-QTST},\lambda}(E)$ according to
\begin{equation}
\big\langle\delta\big(\sigma_a+\sigma_b+E\big)\big\rangle_\lambda=\int_{-\infty}^{\infty}\big\langle\delta\big(\sigma_a+E'+E\big)\delta\big(\sigma_b-E'\big)\big\rangle_\lambda \,\mathrm{d}E', \label{GR-QTST_convolution}
\end{equation} 
where $\sigma_a+\sigma_b=\frac{2}{3}(\mathcal{E}^{(\lambda)}_{0}-\mathcal{E}^{(\lambda)}_{1})$ is just the GR-QTST constraint functional, separated into contributions from the $a$ and $b$ degrees of freedom. Since the $a$ and $b$ degrees of freedom are uncoupled, it follows that the expectation value in the integrand of Eq.~(\ref{GR-QTST_convolution}) can be separated into two terms,
\begin{equation*}
\big\langle\delta\big(\sigma_a+E'+E\big)\delta\big(\sigma_b-E'\big)\big\rangle_\lambda =\big\langle\delta\big(\sigma_a+E'+E\big)\big\rangle_\lambda\big\langle \delta\big(\sigma_b-E'\big)\big\rangle_\lambda.
\end{equation*} 
If we now define
\begin{equation}
p_{a\text{-GR-QTST},\lambda}(E) = \big\langle\delta\big(\sigma_a+E\big)\big\rangle_\lambda, 
\end{equation}
and note that GR-QTST is generally very accurate for systems with a small number of degrees of freedom,\cite{Thapa19,Fang19} we are led to speculate that $p_{a\text{-GR-QTST},\lambda}(0)$ will provide a good approximation to the exact $\rho_\lambda(0)$ and hence give a good approximation to the rate. 

\begin{figure}[t]
 \resizebox{1.0\columnwidth}{!} {\includegraphics{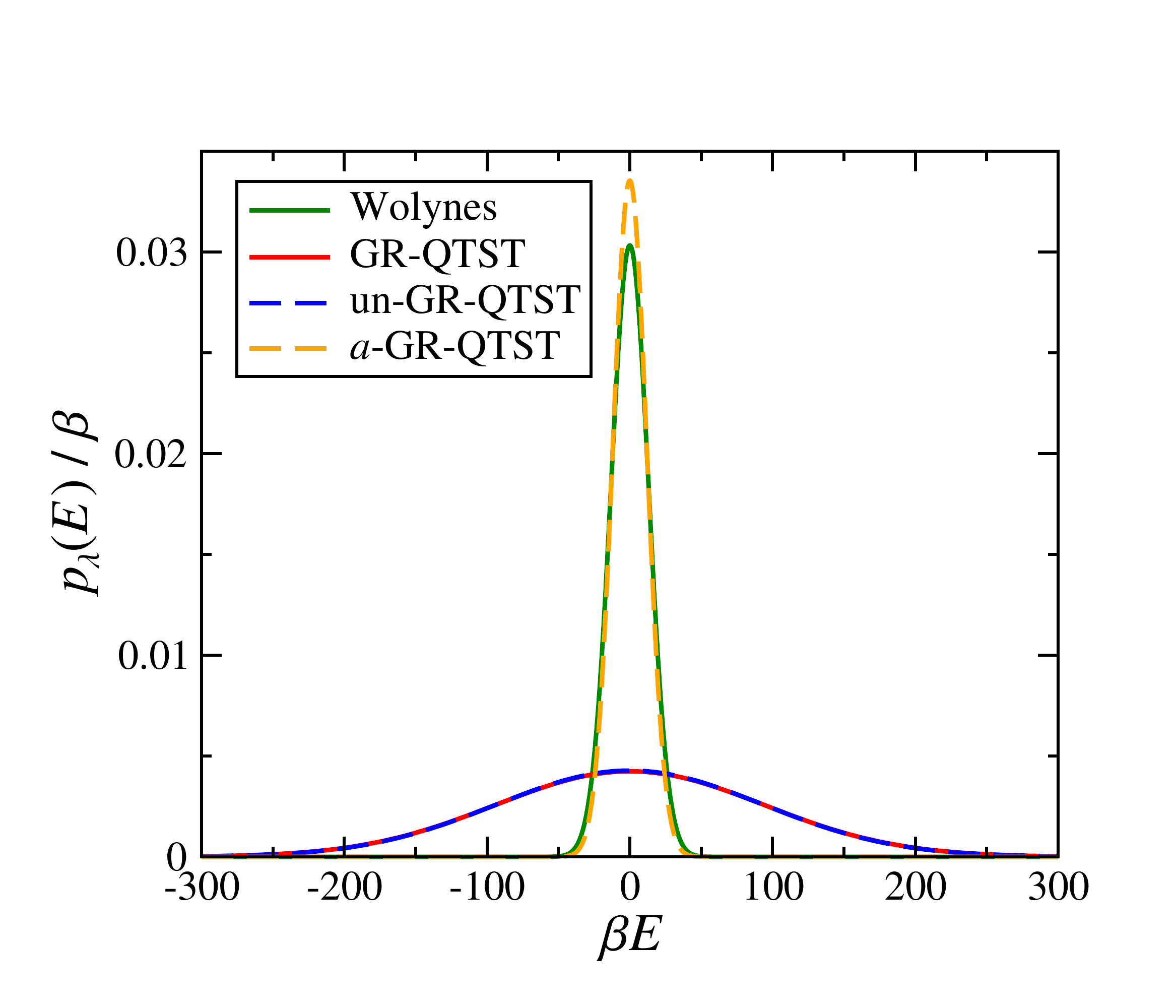}}
 \centering
 \caption{Comparison of the corrected GR-QTST distribution obtained via the approximate deconvolution described in Appendix~C ($a$-GR-QTST), with the raw results for the physical (Fe$^{2+}$-Fe$^{3+}$; GR-QTST) and uncoupled (Fe$^{2.5+}$-Fe$^{2.5+}$; un-GR-QTST) systems, and with the Wolynes distribution.}
 \label{GRQTST_Corrected}
 \end{figure}

In order to extract $p_{a\text{-GR-QTST},\lambda}(E)$ from $p_{\text{GR-QTST},\lambda}(E)$, we can consider the uncoupled $\mathrm{Fe}^{2.5+}\text{-}\mathrm{Fe}^{2.5+}$ system, for which in the present notation
\begin{equation}
V_0(\bm{q}) =V_1(\bm{q}) = \frac{U_{0,a}(\bm{q}_a)+U_{1,a}(\bm{q}_a)}{2} + U_b(\bm{q}_{b}).
\end{equation}
Since we have assumed that there are only a few $a$ degrees of freedom, it follows that the GR-QTST distribution for this uncoupled system will be 
\begin{equation}
p_{\text{un-GR-QTST},\lambda}(E) \simeq \big\langle \delta\big(\sigma_b-E\big)\big\rangle_\lambda.
\end{equation} 
Hence we could in principle numerically deconvolute Eq.~(\ref{GR-QTST_convolution}) to give an approximation to $p_{a\text{-GR-QTST},\lambda}(E)$. However, since we know from the central limit theorem (and from our simulation) that $p_{\text{un-GR-QTST},\lambda}(E)$ will be a Gaussian, and as we have shown that the exact $\rho_{\lambda}(E)$ is approximately a Gaussian, we can reasonably assume that $p_{a\text{-GR-QTST},\lambda}(E)$ will also be Gaussian. Hence, we can simply use the analytical expression for the convolution of two Gaussians to find that 
\begin{equation}
p_{a\text{-GR-QTST},\lambda}(E) \simeq \sqrt{\frac{1}{2\pi\mu_{2,a}^2}}\exp\left(-{E^2\over 2\mu_{2,a}^2}\right),
\end{equation}
where
\begin{equation}
\mu_{2,a}^2 = \int_{-\infty}^{\infty} E^2 \big[p_{\text{GR-QTST},\lambda}(E)-p_{\text{un-GR-QTST},\lambda}(E) \big]\, \mathrm{d}E 
\end{equation}
and we have here made use of the symmetry of the ferrous-ferric system at $\lambda=\lambda_{\mathrm{sp}}=\beta/2$, which implies that the mean of all of the distributions must be zero. Figure \ref{GRQTST_Corrected} shows graphically the result of this approximate deconvolution, illustrating the close agreement between the \emph{a}-GR-QTST results and those of Wolynes theory.


\begin{thebibliography}{100}
\bibitem{Footnote3} For ferrous-ferric electron transfer, the LGR approximation to the first moment of the distribution $\rho_{\lambda}(E)$ is correct at $\lambda=\lambda_{\rm sp}=\beta/2$ by symmetry, but the LGR approximation to the second moment is slightly off.
\end{thebibliography}
\end{document}